\documentclass[sigconf]{acmart}

\AtBeginDocument{%
  \providecommand\BibTeX{{%
    \normalfont B\kern-0.5em{\scshape i\kern-0.25em b}\kern-0.8em\TeX}}}
    
\setcopyright{ccby} 
\copyrightyear{2022}
\acmYear{2022}
\acmConference[CHI '22]{CHI Conference on Human
Factors in Computing Systems}{April 29-May 5, 2022}{New Orleans, LA, USA}
\acmBooktitle{CHI Conference on Human Factors in Computing Systems (CHI '22), April 29-May 5, 2022, New Orleans, LA, USA}
\acmDOI{}
\acmISBN{}

\usepackage{mathtools}

\usepackage{tikz}
\usetikzlibrary{shapes.geometric, arrows,chains, calc, positioning, shapes.multipart,positioning,automata}
\newcommand\vertarrowbox[3][3ex]{%
  \begin{array}[t]{@{}c@{}} #2 \\
  \left\uparrow\vcenter{\hrule height #1}\right.\kern-\nulldelimiterspace\\
  \makebox[0pt]{\scriptsize#3}
  \end{array}%
}

\begin{document}

\title{Passive Haptic Rehearsal for Accelerated Piano Skill Acquisition}

\author{Tan Gemicioglu, Noah Teuscher, Brahmi Dwivedi, Soobin Park, Emerson Miller}
\email{{tgemici,nteuscher3,bdwivedi6,spark878,emiller88}@gatech.edu}
\affiliation{%
  \institution{Georgia Institute of Technology}
  \city{Atlanta, GA}
  \country{USA}}
  
  



\author{Celeste Mason}
\email{celeste.mason@uni-hamburg.de}
\affiliation{%
  \institution{University of Hamburg}
  \city{Hamburg}
  \country{Germany}}

\author{Caitlyn Seim}
\email{cseim@stanford.edu}
\affiliation{%
  \institution{Stanford University}
  \city{Stanford, California}
  \country{USA}}

\author{Thad Starner}
\email{thad@gatech.edu}
\affiliation{%
  \institution{Georgia Institute of Technology}
  \city{Atlanta, GA}
  \country{USA}}
\renewcommand{\shortauthors}{Gemicioglu et al.}

\begin{abstract}
Passive haptic learning (PHL) uses vibrotactile stimulation to train piano songs using repetition, even when the recipient of stimulation is focused on other tasks. However, many of the benefits of playing piano cannot be acquired without actively playing the instrument. In this position paper, we posit that passive haptic rehearsal, where active piano practice is assisted by separate sessions of passive stimulation, is of greater everyday use than solely PHL. We propose a study to examine the effects of passive haptic rehearsal for self-paced piano learners and consider how to incorporate passive rehearsal into everyday practice.
\end{abstract}



\captionsetup[figure]{skip=-2pt}
\begin{teaserfigure}
  \vspace{-5mm}
  \includegraphics[width=\textwidth]{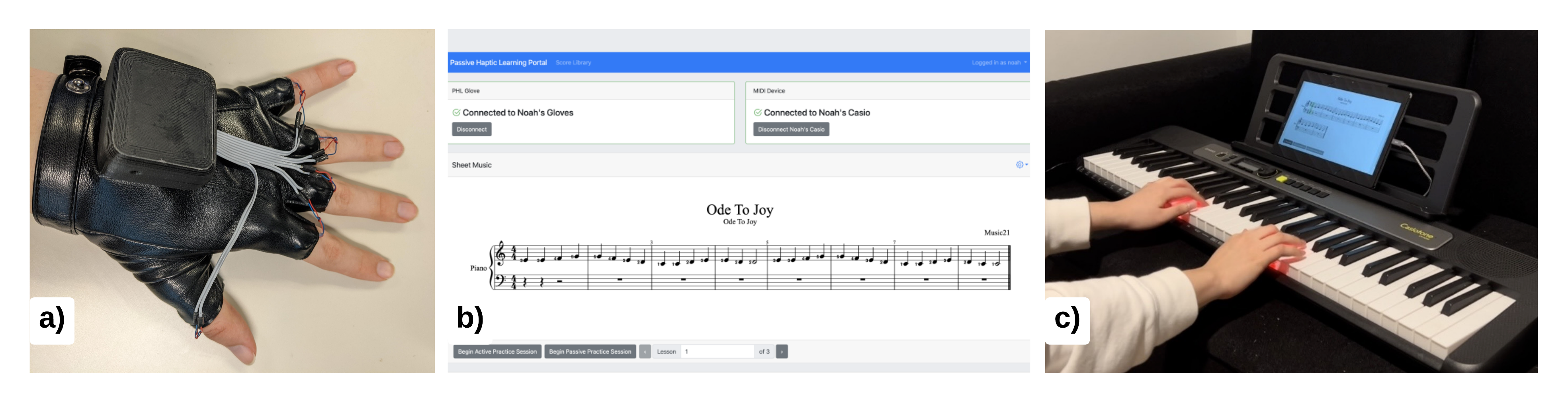}
  \caption{(a) Passive haptic rehearsal gloves. (b) Web app for active and passive practice. (c) Active practice on light-up keyboard.}
  \Description{Teaser image featuring passive haptic rehearsal gloves, web app with sheet notes displayed, and person playing light-up keyboard for active session.}
  \label{fig:teaser}
\end{teaserfigure}
\maketitle

\section{Introduction}

Playing piano improves mental health and has a positive impact on welfare \cite{mccaffrey_music_2016}. However, the learning process can take significant time between learning theory and technique, practicing new piano pieces and rehearsing previously learned ones. Passive haptic learning (PHL) offers an alternative method of rehearsal, speeding the learning process and increasing retention \cite{donchev_investigating_2021}.

Previous PHL studies have involved almost entirely passive training, with users only actively playing on a keyboard for evaluation \cite{seim_towards_2015}. This research prioritized internal validity, focusing on narrow conditions: simple, short songs, rigid pedagogical methods, and inexperienced students. If PHL is to be a successful piano education tool, it must be usable by real piano students learning complicated songs and engaging in multiple forms of practice through various mediums. We expand upon existing work by integrating passive training into a piano learning program we hope to be suitable for songs and students of all levels. We propose passive haptic rehearsal, in which students learn songs through a combination of active practice and passive training sessions, to reinforce developing skills. 



\begin{figure*}
\includegraphics[width=\textwidth]{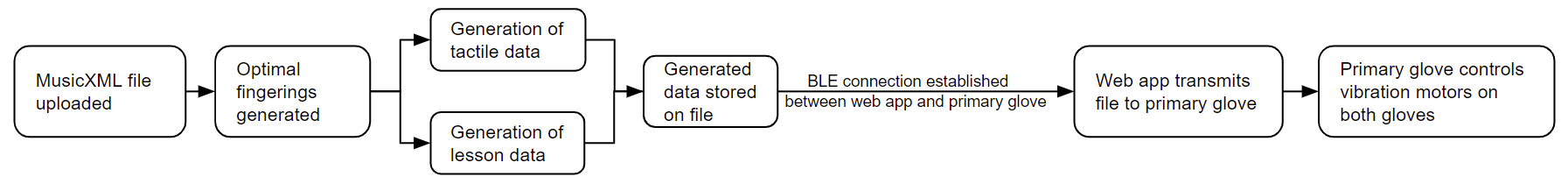}
\caption{Deriving vibrotactile stimulation from sheet music} \label{fig:M2}
\end{figure*} 

\section{Related Work}

Existing work in piano skill acquisition primarily focuses on learning assisted by a piano instructor. We focus instead on self-paced piano learning where the user learns without any direct human assistance. Consistent feedback can encourage learners' progress \cite{costa-giomi_piano_2005}, and the process can be gamified by providing more incentives as the learner spends time practicing piano, ensuring continuity and regular practice \cite{raymaekers_game_2014}. New piano interfaces seek to speed learning and reduce the cognitive load demanded by practice. Visual cues, such as light-up keys, new piano notations \cite{rogers_pino_2014} and finger position tracking \cite{marky_lets_2021} have enabled such advances in active learning. 

Passive learning, in contrast, is "caught, rather than taught" \cite{krugman_passive_1970} and can enable learning even in the absence of motivation and effort \cite{zukin_passive_1984}. PHL uses vibrotactile stimulation to teach piano passively using intensive repetition of instructional cues applied directly to the fingers \cite{huang_pianotouch_2008}. Surprisingly, learning occurs whether or not the tactile stimulation is accompanied by audio of the notes playing \cite{markow2012mobile}. Early studies extensively compared passive and active learning for piano and confirmed that PHL did not occupy the user's attention during learning \cite{kohlsdorf_mobile_2010,huang_mobile_2010}.
Newer research has expanded from simple melodies to complete, two-handed chorded piano pieces \cite{seim_towards_2015}. Furthermore, recent work shows that passively learned note sequences can be recalled with greater accuracy than actively learned sequences when recall is supported by audiovisual cues \cite{donchev_investigating_2021}.

\section{Passive Piano Learning Framework}



Our proposed framework includes a pair of vibrotactile haptic feedback gloves, a Bluetooth-capable tablet such as a Samsung Galaxy Tab A8 and a Casio LK-S250 keyboard. The Casio keyboard can illustrate how to play a song using light-up keys and can capture the user's practice sessions using MIDI. We use soft, faux leather driving gloves with sizes varying for each user. The PHL gloves contain five vibration motors (Precision Microdrives 310-103) at the base of the fingers and a 3D-printed case at the back of the palm to protect the electronics from sweat, dust and other damage.

Our web app allows users to manage and interact with scores and the selected hardware.  Users can interact with a song in a number of ways, such as viewing the sheet music, playing through the song with audio and visual cues, recording their own performances of a song, and taking personal practice notes. All users also have access to songs in a public corpus (e.g. the study curriculum). Users can also upload their own songs as MusicXML files, which are processed and stored in their account. Our software makes heavy usage of the computational musicology library \emph{music21} for processing and interacting with digital scores \cite{cuthbert2010music21}.

When new songs are uploaded to our web app, they are processed in two significant ways. First, we automatically produce fingering data for each note in the song, which is necessary for tactile vibrations but not contained in most MusicXML scores. We utilize the open source \emph{pianoplayer} package to produce optimal fingering patterns \cite{musy_pianoplayer_2022}. Second, because passive training works optimally on short segments of 10-17 stimuli, we chunk each song into small ``lessons,'' which are repeated multiple times during passive training. This ``chunking'' follows the guidelines established by Seim \cite{seim_wearable_2019}.

The web app allows users to connect the tablet to the PHL gloves over Bluetooth LE or to the keyboard over USB. When connected to the gloves, they can initiate passive training sessions with a variety of parameters such as length of session and number of lesson repeats. When connected to the keyboard, they can record and upload their performance of a song, which is then evaluated by our scoring algorithm (discussed below). 
\\
\subsection{Integrating PHL with Active Practice}

We restrict active piano lessons to 30 minutes to prevent fatigue, though students are allowed to use whatever active practice techniques they prefer.
We incorporate feedback as an incentive and a method of gamifying the learning process. First, the Casio keyboard demonstrates the song to be learned by playing the song and lighting the keys in synchrony. The learner attempts to repeat the score and then proceeds to their active lesson. Afterwards, the keyboard again demonstrates the song, and the learner tries to repeat it. The web app then graphs the student's progress. Feedback immediately after, but not during, the lesson allows students to learn from their mistakes faster and build confidence as they remedy errors. Offering feedback gives the participant a sense of accomplishment and a frame of reference over time, thus encouraging progress \cite{costa-giomi_piano_2005}.

Dual cuing has been shown to greatly improve performance in active learning \cite{katz_optimizing_2021}. During active practice, we display the piano keys and sheet music on the web app while concurrently playing the song. This process increases learning and reduces frustration \cite{seim_towards_2015}.

In addition to the 30 minute active practice session each day, learners wear the PHL gloves for 2.5 hours. Using vibration motors at the base of each finger, the gloves repeatedly stimulate the user's fingers in the order and rhythm in which the piano keys are played.  The gloves are fingerless so as not to interfere with the user's everyday tasks. Since no audio is played and the stimulation is relatively subtle, the user can ignore the gloves. The gloves are self-contained in that they do not need a continuous connection to an external device to operate. The right glove controls the left through Bluetooth LE. Battery life is around three hours.

Testing the learner before and after active sessions also provides data on passive rehearsal. The test at the end of the previous active session provides a baseline to compare to the test at the beginning of the current active session to determine how much ``forgetting'' has happened between sessions. Without passive stimulation, we expect significant degradation of performance, but in pilot testing with passive rehearsal, learners seem to retain or even improve upon the benefits of their previous active learning session \cite{huang_pianotouch_2008}.

\subsection{Performance Evaluation}

To evaluate the performance of a user, we treat the notes played as a temporal sequence $P$ for calculating $Eval(P)$. We use the Needleman–Wunsch algorithm \cite{Needleman1970AGM}, a dynamic programming algorithm for optimal global sequence alignment, for matching the note sequence played to the note sequence in the original piece. Previous work utilized Dynamic Time Warping for sequence alignment which ignores the error of multiple presses of the same key (the error between the key-press sequences “C A A B” and “C A B” is zero) \cite{huang_mobile_2010,seim_towards_2015}. Needleman-Wunsch is ``stricter'' when assigning error costs. In addition to alignment errors, we have included a timing error in our new algorithm to measure rhythmic inaccuracies. Once the two sequences are matched, a timing error is applied to matched key presses in which the time difference between user played and original key presses crosses a pre-determined threshold. 
\\

\begin{tabular}{c c c c c c c c c c}
 Original Times: & $t_1$ & $t_2$ & - & $t_3$ & $t_4$& - & $t_5$ & $t_6$ & - \\ 
 Original Notes:&  C & B & B & A & E& E & F & C & - \\  
 Performed Notes: &  C & B & - & A & E& C & F & C &  C\\
   Performed Times: & $t_1'$ & $t_2'$ & \vertarrowbox{\text{-}}{Deletion} & $t_3'$ & $t_4'$& \vertarrowbox{\text{-}}{Substitution} & $t_5'$ & $t_6'$ & \vertarrowbox{\text{-}}{Insertion} 
\end{tabular}
\\For the above example, assuming $m$ total matched notes and the timing error threshold to be $T$:
$$\text{Timing Cost = Timing cost of matched notes = } \sum_{i=0}^m 1(|t_i - t_i'|\geq T)$$
$$\text{Alignment Cost = Deletion cost + Insertion cost + Substitution cost} $$$$= 1+1+1 = 3$$
Alignment and timing can be weighted to give an overall evaluation of the performance; with weights $W_a$ and $W_t$ as the weights of alignment cost and timing cost respectively:
$$\text{Total Cost }=  W_a*\text{Alignment Cost} + W_t* \text{Timing Cost} $$
Key presses that occur in rapid succession are extracted as chords and treated as unordered sequences.

\section{Proposed Study}


Since we focus on self-paced piano learning, we wish to recruit hobbyist piano learners who are practicing in their own time without formal piano lessons. We plan to reach these participants by advertising to people who watch content from online piano instructors on video streaming platforms. 16 participants will be recruited and paired by skill level based on the improvements they achieve while practicing a previously unseen piece actively for half an hour.


For the study, we chose a pair of songs from the composer Bartok: Children at Play and Young Men’s Dance. Our participants likely have not heard or learned these songs before, and a disqualification criteria during recruitment  will be if they have learned these songs previously. The songs are of a difficulty level and length such that they are not trivial to learn but our target subjects should be able to adequately learn them in a two week period.


The study will follow a counter-balanced Latin Square design for four conditions: two each for starting one of two piano pieces and starting with the sham or functional gloves. Over the course of 2 weeks, participants will be required to log a total of 3 hours of practice per day: 30 minutes of active practice and 2.5 hours of passive practice. 

During active practice, subjects have access to our interactive web app, which allows them to view sheet music, play back songs with audio and visual cues and record their own performances. However, we will not mandate particular modes of active practice (i.e. scales, repetitions, metronome work, etc). During passive practice, the subjects will wear our PHL gloves, and may go about regular daily tasks during the session. 

We will bring subjects into the lab three times throughout the study for more in-depth evaluations: once at the beginning of the study, once at the midway point (two weeks), and once at the conclusion. These evaluations will consist of recording their unaided performance of each of our curriculum songs.  Likert scale and preference surveys and free responses about their progress will also be compared between the conditions. 
 
The experiment is designed to be both within-subjects and between subjects. Each pair will be assigned the same song. One learner will be using a fully operational glove while the other will be given a similar glove and be told that the vibrations are set at a level below perception (when in fact they will not be vibrating at all). Using the daily pre and post active session tests, we will compare learning and ``forgetting'' rates between subjects. After two weeks, the pair will change to the second song, and we will switch the sham and active glove condition. In this manner we can compare each participant's performance to themselves between the sham and active condition. Note that, because of the pairing, participants must be recruited and randomized in teams of eight to maintain counterbalancing. ANOVAs and permutation tests will be used to determine if the learning rates differ statistically. 


\begin{figure} \includegraphics[width=80mm,scale=1]{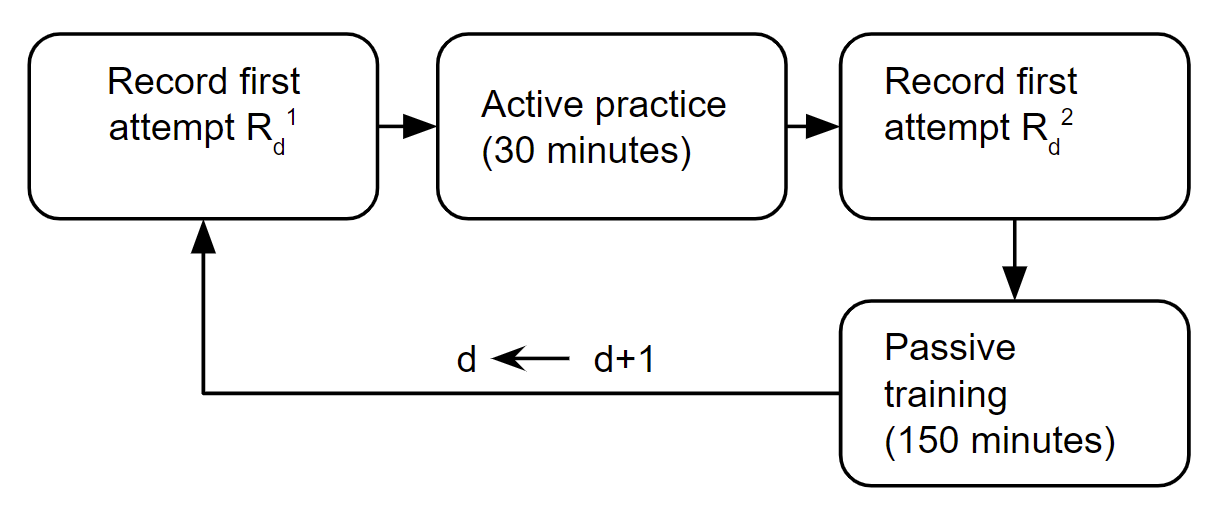}
\caption{Sessions scheduled on day $d$, where we measure \\progress from active practice as $Eval(R_d^2) - Eval(R_d^1)$ and progress from passive rehearsal as $Eval(R_{d+1}^1) - Eval(R_d^2)$} \label{fig:M1}
\end{figure} 

\section{Conclusion}

This work presents a new framework for PHL by designing a curriculum that combines active learning with passive haptic rehearsal. We focus our efforts on hobbyist piano learners as they may most appreciate the benefits of the PHL gloves. In order to examine the effects of passive haptic rehearsal, we propose a controlled, randomized study to test the augmented piano learning process in the users’ own environments over the course of four weeks. 


\section{Acknowledgments}

This research was supported by the National Science Foundation, Partnerships for Innovation grant \#2122797 and the Georgia Research Alliance, Inc. based in Atlanta, Georgia.

\bibliographystyle{ACM-Reference-Format}
\bibliography{sample-sigconf}


\end{document}